\documentclass[12pt,thmsa,a4paper]{article}

\usepackage{amsmath}
\usepackage{amssymb}

\newcommand{\plabel}{\label}

\newcommand{\beqs}{\begin{equation*}}
\newcommand{\beq}{\begin{equation}}

\newcommand{\eeqs}{\end{equation*}}     
\newcommand{\eeq}{\end{equation}}

\newcommand{\beqas}{\begin{eqnarray*}}
\newcommand{\beqa}{\begin{eqnarray}}

\newcommand{\eeqas}{\end{eqnarray*}}
\newcommand{\eeqa}{\end{eqnarray}}

\newcommand{\eq}[2]{\begin{equation} #1 \plabel{#2} \end{equation}}

\newcommand{\al}{\alpha}
\newcommand{\be}{\beta}

\newcommand{\Om}{\Omega}

\newcommand{\blist}{\begin{itemize}}

\newcommand{\elist}{\end{itemize}}

\begin{document}

\begin{titlepage}
\renewcommand{\thefootnote}{\fnsymbol{footnote}}

\hfill TUW--00--32 \\

\begin{center}

{\Large\bf 2D gravity without test particles is pointless (Comment on 
hep-th/0011136)}\\
  \vspace{7ex}
  D.~Grumiller\footnotemark[1],
  D.~Hofmann\footnotemark[2],
  W.\ Kummer\footnotemark[3],
  \vspace{7ex}

  {\footnotemark[1]\footnotemark[2]\footnotemark[3]\footnotesize Institut f\"ur
    Theoretische Physik \\ Technische Universit\"at Wien \\ Wiedner
    Hauptstr.  8--10, A-1040 Wien, Austria}
  \vspace{2ex}

   \footnotetext[1]{E-mail: \texttt{grumil@hep.itp.tuwien.ac.at}}
   \footnotetext[2]{E-mail: \texttt{hofmann@hep.itp.tuwien.ac.at}}
   \footnotetext[3]{E-mail: \texttt{wkummer@tph.tuwien.ac.at}}
\end{center}
\vspace{7ex}
\begin{abstract}
Claims of a general Weyl invariance of an arbitrary 2D dilaton theory are
critically discussed. 
\end{abstract}

\end{titlepage}

\section{Introduction}

The explicit or implicit assumption that generic (matterless) dilaton 
theories in two dimensions which are related by Weyl (conformal) 
transformations lead to ``equivalent'' formulations, has quite a 
long history \cite{bal91}. A recent note \cite{cav00b} raises this issue again
and contains statements which require critical comment.

Conformal transformations are very useful in many contexts of physics, ranging 
from classical electrodynamics to string theory. They are especially 
convenient in particular 2D models, where the absence of an extrinsic scale 
implies invariance under {\em local} non-singular conformal 
transformations (cf. e.g. \cite{fms97}). 
This seems to be the reason why they have also attracted attention of the 
community studying 2D dilaton theories of the form\footnote{We use the 
same notation as in \cite{grk00}.}
\eq{
S=\int_{M_2} \sqrt{-g}\left[XR-U(X)\left(\nabla X\right)^2 + V(X)\right].
}{i1}
In particular, the field-dependent Weyl transformation
\eq{
g_{\al\be} = \Om(X)^{-2} \tilde{g}_{\al\be}, \hspace{0.5cm}
\Om(X) = \exp{\left[-\frac{1}{2} \int^X U(X')dX'\right]}
}{i2}
has been used to simplify (\ref{i1}) to
\eq{
\tilde{S}=\int_{\tilde{M}_2} \sqrt{-\tilde{g}}\left[X\tilde{R} + \tilde{V}(X)\right]
}{i3}
with $V(X) = \Om^2 \tilde{V}(X)$. 

However, it should be stressed that for a 
large class of models\footnote{E.g. theories with 
$U(X) = aX^{-1}$ with $a \in \mathbb{R}$, including spherically 
reduced gravity (SRG) and the CGHS model \cite{cgh92}.} $U(X)$ in (\ref{i1})
and hence the conformal factor (\ref{i2}) are singular at the ``origin'' 
(where the curvature singularity is ``located''), because $\lim_{X \to 0} 
\Om(X) = 0$. In addition, on dimensional grounds the last term in (\ref{i1})
must contain a scale. Moreover, the only invariance transformations of 
(\ref{i1}) are diffeomorphisms. A nontrivial redefinition of the potentials
$U(X)$ and $V(X)$ goes beyond the usual definition of an invariance. 
Additionally, the dilaton may carry a conformal weight whenever the 2D model
(\ref{i1}) stems from dimensional reduction \cite{ghk00}.

These simple observations show that even at the classical level serious
problems are likely to occur when the singularity of that transformation
and the scale dependence are not duly taken into account.
Indeed this is confirmed by consideration of the geodesics of test particles 
and their consequences for the causal structure of spacetime.

Since all the arguments given below appeared already in several papers 
(reaching back at least half a century \cite{fie56}) we restrict ourselves 
to a brief qualitative discussion and refer to the literature for 
a more detailed analysis.

\section{Classical observables}

One of the key ingredients of the attempt to prove Weyl invariance in 
\cite{cav00b} is the premise that the only physical observable is the 
conserved quantity, which is present in all such theories 
\cite{kus92}. It is true that the conserved 
quantity ${\cal C}$ is proportional to the ADM mass $M$ in a given 
conformal frame. But conformal invariance of ${\cal C}$ does not imply 
automatically conformal invariance of $M$. Any (local) function of 
${\cal C}$ will be again a conserved quantity and there is no preferred 
way to relate ${\cal C}$ with $M$ in different conformal
frames. Thus, the physical observable $M$ is in general not invariant under
conformal transformations. This has been exploited in detail in 
\cite{alv97}. We claim that 
also the scalar curvature is a classically accessible physical observable, 
which can be obtained by investigating the geodesics of lightlike and timelike 
test particles (i.e. no backreactions are involved). Indeed, many textbooks
about general relativity use the geodesic deviation equations to motivate the
concept of curvature (cf. e.g. \cite{wal84}). Moreover, every time a 
conformal diagram is constructed in order to discuss the causal structure the 
geodesics of test particles are used (at least implicitly). It is a 
non-negligible difference whether they reach the singularity with finite 
affine parameter or not. Thus geometry without test particles 
to probe it has no well-defined meaning. 

The r{\^o}le of geometric variables 
in gravity theories is twofold: On the one hand they represent fields, 
analogous to gauge fields on a fixed background. On the other hand, 
$g_{\mu\nu}$ is identified as the metric of the twodimensional manifold, which
is exploited by the geodesics of test particles calculated from {\it that}
$g_{\mu\nu}$.
As noted correctly in the conclusions of \cite{cav00b} explicit coupling to
matter fields breaks Weyl invariance, in general. But already the (at least 
implicit) inevitable presence of test particles has the {\em same} 
consequence, although they usually are not regarded as ``matter'' because they 
have no influence on the metric by assumption.  

Even if the action and the equations of motion were Weyl
invariant, the geodesics of test particles are not. Since we 
regard the causal structure of spacetime and the Ricci 
scalar as geometrical properties (although one needs test particles to probe 
them), we conclude that the geometry itself is {\em not} Weyl invariant.

In fact, this discussion has quite a long history. Already Fierz pointed out
that after performing a Weyl transformation one has to {\em transform in 
addition the geodesics of test particles} and that they no longer obey
the equation for geodesics calculated with the metric of the transformed 
geometry \cite{fie56}. This observation was also made in the 
original work of Jordan \cite{jor55} and Brans and Dicke \cite{brd61} who were 
the first to consider scalar-tensor theories. 

\section{Quantum observables}

At the quantum level the situation becomes even worse.
The field quantization brings in a natural scale, the vacuum energy,
that breaks explicitly conformal invariance in those special cases where the
classical theory is conformally invariant.
Not surprisingly, there is also ample evidence that the 
flux of Hawking radiation
depends on the choice of the conformal frame \cite{cfn99}.
This is to be expected because the asymptotic flux is measured at infinity
and hence it is a global property that can be changed under a conformal 
transformation. The issue of Hawking radiation in 2D dilaton 
theories is not settled completely, although considerable progress in 
calculating the Hawking flux including backreactions from the conformal 
anomaly of a scalar field \cite{MWZ}
and a dilaton anomaly has been achieved \cite{kuv99} 
(see also references therein).

Finally, we should stress that within the path integral approach to 2D quantum
gravity \cite{klv99} a field redefinition (\ref{i2}) introduces functional 
determinants with unmanagable problems  following from the inevitable 
singularities present in such a transformation.

\section*{Acknowledgement} 
We thank D.V. Vassilevich for discussions and
careful reading of the manuscript. This work has been supported by Project 
P14650-TPH of the Austrian Science Foundation (\"Osterreichischer Fonds zur 
F\"orderung der wis\-sen\-schaft\-lichen Forschung).

\end{document}